\documentclass[apj]{emulateapj}

\usepackage{amsmath,amssymb,verbatim} 
\usepackage{color} 
\usepackage{natbib}

\shorttitle{Intermittent Pulsars}
\shortauthors{}

\begin{document}
\title{On the Spin-Down of Intermittent Pulsars}


\author{Jason Li$^1$, Anatoly Spitkovsky$^1$, \& Alexander
Tchekhovskoy$^2$} \affil{$^1$Department of Astrophysical Sciences, Peyton Hall,
Princeton University, Princeton, NJ 08544, USA\\ $^2$Princeton Center for
Theoretical Science, Jadwin Hall, Princeton University, Princeton, NJ
08544, USA} 
\email{jgli@astro.princeton.edu}

\begin{abstract}
Magnetospheres of pulsars are thought to be filled with plasma, and
variations in plasma supply can affect both pulsar emission properties
and spin-down rates.  A number of recently discovered ``intermittent''
pulsars switch between two distinct states: an ``on'', radio-loud
state, and an ``off'', radio-quiet state.  Spin-down rates in the two
states differ by a large factor, $\sim 1.5-2.5$, which is not easily
understood in the context of current models.  In this Letter we
present self-consistent numerical solutions of ``on'' and ``off''
states of intermittent pulsar magnetospheres.  We model the ``on''
state as a nearly ideal force-free magnetosphere with abundant
magnetospheric plasma supply.  The lack of radio emission in the
``off'' state is associated with plasma supply disruption that results
in lower plasma density on the open field lines. We model the ``off''
state using nearly vacuum conditions on the open field lines and
nearly ideal force-free conditions on the closed field lines, where
plasma can remain trapped even in the absence of pair production.  The
toroidal advection of plasma in the closed zone in the ``off'' state
causes spin-downs that are a factor of $\sim 2$ higher than vacuum
values, and we naturally obtain a range of spin-down ratios between
the ``on'' and ``off'' states, $\sim 1.2-2.9$, which corresponds to a
likely range of pulsar inclination angles of $30{-}90^\circ$.  We
consider the implications of our model to a number of poorly
understood but possibly related pulsar phenomena, including nulling,
timing noise, and rotating radio transients.

\end{abstract}



\section{Introduction}\label{sec:intro}

Pulsars spin down due to torques exerted by currents flowing on the
surface of the neutron star.  In the absence of magnetospheric plasma,
the pulsar spins down due to magneto-dipole radiation
\citep{Michel91,Beskin93}.  Magnetospheric plasma allows for currents
that can produce additional spin-down torques \citep[][hereafter
S06]{Spitkovsky06}, so variations in plasma supply can potentially
modulate pulsar spin-down.

Several classes of pulsars have been identified that may exhibit such
modulation.  Nulling pulsars have radio emission that appears to shut
off for a few to several tens of rotation periods.  Since radio
emission is presumably tied to the magnetospheric plasma, the nulling
suggests that some process is affecting the plasma supply and/or
currents above the polar caps \citep{Wang07,Zhang07,Timokhin10}.
Intermittent pulsars switch between an ``on'', radio-loud, state in
which they behave like normal radio pulsars, and an ``off'',
radio-quiet, state in which they produce no detectable radio emission
for long periods of time.  This process may be an extreme
manifestation of nulling.  The first two intermittent pulsars with
published data have quite different duty cycles: PSR B1931+24
\citep[][hereafter K06]{Kramer06}, with a period $P\approx0.8$~s,
cycled through the ``on''--''off'' sequence of states approximately
once a month, whereas PSR J1832+0029 \citep{Lyne09}, with a period
$P\approx0.5$~s, kept quiet for nearly two years between ``on''
cycles of unknown length.  The spin-down rate for each of these
pulsars is larger in the ``on'' state than in the ``off'' state by a
factor $f_{{\rm on}\to{\rm off}}\simeq1.5$.  A third intermittent
pulsar, PSR J1841-0500, was not detected for over $1.5$ years between ``on''
cycles, one of which appears to have lasted for at least a year
\citep{Camilo11}.  This pulsar seems to have a spin-down ratio between
``on'' and ``off'' states of $f_{{\rm on}\to{\rm off}}\simeq2.5$.
Such substantial differences in spin-down rates suggest that the
pulsar magnetosphere undergoes a dramatic reconfiguration as it
transitions between the ``on'' and ``off'' states, yet such a
transition was reported for PSR B1931+24 to take place in just over
$10$ pulsar periods.

Intermittent pulsars offer a unique testbed of pulsar theory.  K06
first proposed that in the ``on'' state plasma fills the pulsar
magnetosphere and supports plasma processes that produce radio
emission.  The pulsar transitions to the ``off'' state when open field
lines become depleted of charged radiating particles.  K06
approximated the spin-down rate in the ``off'' state by the spin-down
of a vacuum dipole and estimated the extra plasma currents needed to
account for the observed spin-down of the ``on" state. It is not
clear, however, whether a working pulsar should naturally yield the
required on-off spin-down ratio of this picture.  The simplest model
for the ``on'' state is the force-free magnetosphere, which has
abundant charges everywhere.  The force-free spin-down rate is larger
than the vacuum spin-down rate by a factor $f_{{\rm ff}\to{\rm
vac}}=(1+\sin^2\alpha)/(2/3 \sin^2\alpha)$ (S06) that is greater than
or equal to $3$ for all inclination angles $\alpha$.  This is clearly
incompatible with the observed values, $f_{{\rm on}\to{\rm
off}}\simeq1.5-2.5$ \citep[][]{BN07,Gurevich07}. This suggests that,
perhaps, we do not understand the spin-down power of the ``off" state.

The shutoff of pair formation as the intermittent pulsar switches
``off'' allows plasma to escape along the open field lines, but the
plasma in the closed zone is confined by the geometry of the field
lines.  This is an important physical effect that was not included in
previous work (K06; \citealt[][hereafter
LST11]{Li11};\citealt{kalap11}).  The currents and charges associated
with plasma trapped in the closed zone can increase the spin-down in
the ``off'' state even if the open field lines are empty.  Thus, in
this paper, we model the ``off'' state using a simple two-zone
prescription in which the closed zone is highly conducting and the
open field lines are vacuum-like.  We run resistive force-free
simulations to directly solve for the magnetospheric geometry in the
``on'' and ``off'' states and test whether this model can produce the
observed intermittent pulsar spin-down ratios.  In Section
\ref{sec:setup} we describe the numerical code and setup.  Section
\ref{sec:results} illustrates our intermittent pulsar solutions and
shows the spin-down results.  Section \ref{sec:discussion} provides a
brief summary of our results and their observational implications.

\section{Setup}\label{sec:setup}
We employ a three-dimensional numerical code (see S06) that implements
the finite difference time-domain scheme
\citep[FDTD,][]{TafloveHagness05} to evolve electromagnetic fields
from Maxwell's equations,
\begin{align}
\label{maxwell}
\partial_t \vec{E} &=c \vec{\nabla}\times \vec{B} - 4\pi\vec{j},  \nonumber \\
\partial_t \vec{B} &= -c\vec{\nabla}\times \vec{E},
\end{align}
where the current $\vec{j}$ is given by 
\begin{equation} 
\vec{j}=\rho\vec{v}+\sigma \vec{E}_{\rm fluid}.
\label{eq:current}
\end{equation}
The fluid velocity $\vec{v}=c(\vec{E}\times\vec{B})/(B^2+E_0^2)$ is
the generalized drift velocity, $\vec{E}_{\rm
fluid}=\gamma(\vec{E}+\vec{v}\times\vec{B})$ is the fluid frame
electric field, $\gamma=(1-v^2/c^2)^{-1/2}$, $E_0$ is the magnitude of
$\vec{E}_{\rm fluid}$ (see LST11), $\rho$ is the charge density, and
$\sigma$ is the plasma conductivity in the fluid frame.  The central
region of our grid is occupied by a conducting spherical star of
radius $R_*$, rotating at angular velocity $\vec{\Omega}$, with
embedded dipole field of magnetic moment $\vec{\mu}$ inclined relative
to the rotation axis by angle $\alpha$.  We resolve the light cylinder
$R_{\rm LC}=c/\Omega$ with 80 cells and set $R_*=30$ cells.  See LST11
for a detailed description of our code and resistive current
formulation.  We have verified that our solutions are converged with
spatial resolution, as well as run sufficiently long so as to reach a
steady state in the frame corotating with the pulsar.

In LST11 we showed that our resistive force-free formulation can
capture both the vacuum and ideal force-free limits by varying the
conductivity parameter.  We model the ``on'' state as a magnetosphere
with high conductivity $(\sigma/\Omega)^2=40$, our fiducial value
representing force-free--like conditions for $R_*/R_{\rm LC}=3/8$.
This is preferable to using an ideal force-free formulation as in S06,
because our high conductivity solutions are numerically cleaner,
especially in the current sheets (LST11).  We model the ``off'' state
as having conducting and vacuum-like regions separated sharply at the
boundary of the closed field line region, which we approximate as the
closed field lines of the force-free ``on'' state.  Nonrotating dipole
magnetic field lines are traced by the curves $s=s_0\sin^2\theta$
\citep{MichelLi99}, where $\theta$ is the angle from the magnetic
axis, $s$ is spherical radius, and $s_0$, the maximum perpendicular
distance of a field line from the magnetic axis, specifies which field
line is under consideration.  The closed field lines in the force-free
solutions are typically stretched in the direction perpendicular to
the magnetic axis as compared to vacuum dipole closed field lines.  We
use the surface $s=R_{\rm LC}\sin^2\theta(R_{\rm
LC}\sin^2\theta/R_*)^k$ to demarcate the closed field line region.
The exponent $k$ is set at each inclination angle to best match the
shape of the force-free closed zone (see Section \ref{sec:results} for
illustrations of these demarcation surfaces), and the half-angle size
of the conducting polar cap is given by $\theta_{\rm
pc}=\sqrt{R_*/R_{\rm LC}}$.  Interior to the demarcation surface, we
set $(\sigma/\Omega)^2=40$, as in the ``on'' state.  Exterior to the
demarcation surface, we set $(\sigma/\Omega)^2=0.04$, a fiducial value
representative of vacuum-like conditions for $R_*/R_{\rm LC}=3/8$.  We
pick this value over $\sigma/\Omega=0$ for better numerical accuracy
when computing the field geometry.  This choice of conductivity on
open field lines and the exact shape of the demarcation surface have
minimal effect on spin-down (see Section \ref{sec:results}).

\section{Results}\label{sec:results}

\begin{figure*}[htp]
\centering
\includegraphics[width=0.5\textwidth]{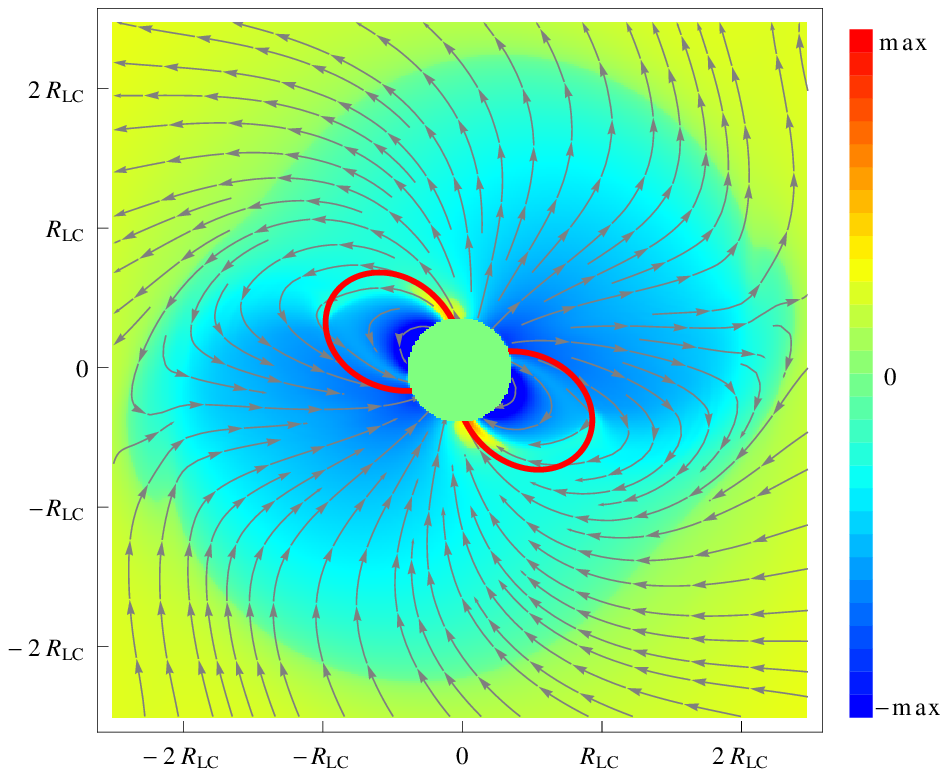}\hfill
\includegraphics[width=0.5\textwidth]{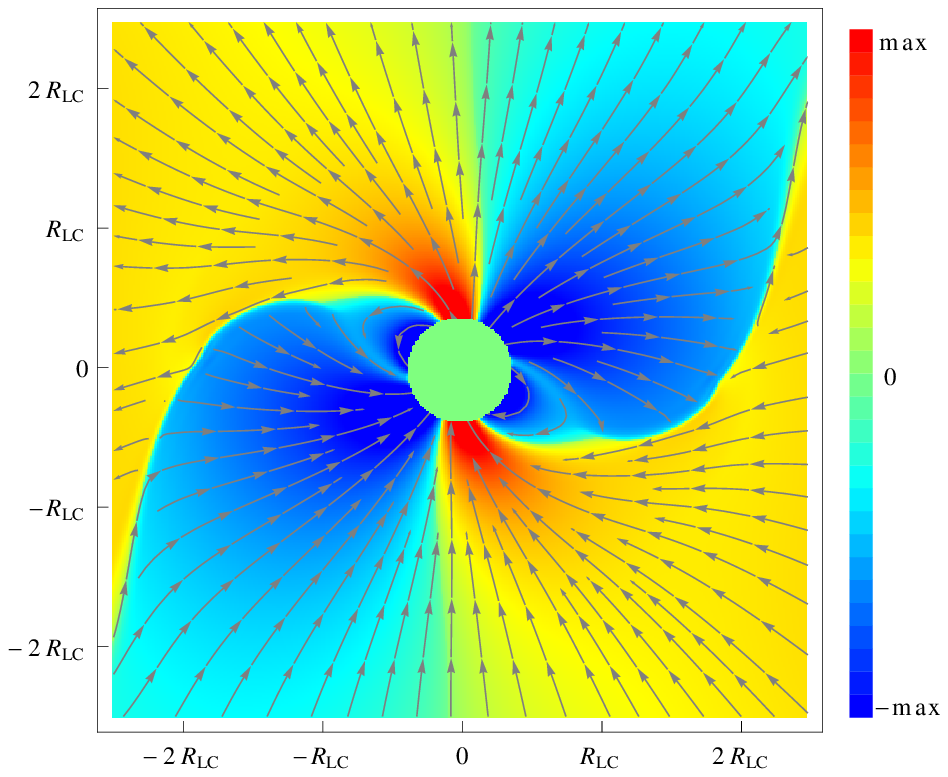}\\
\includegraphics[width=0.5\textwidth]{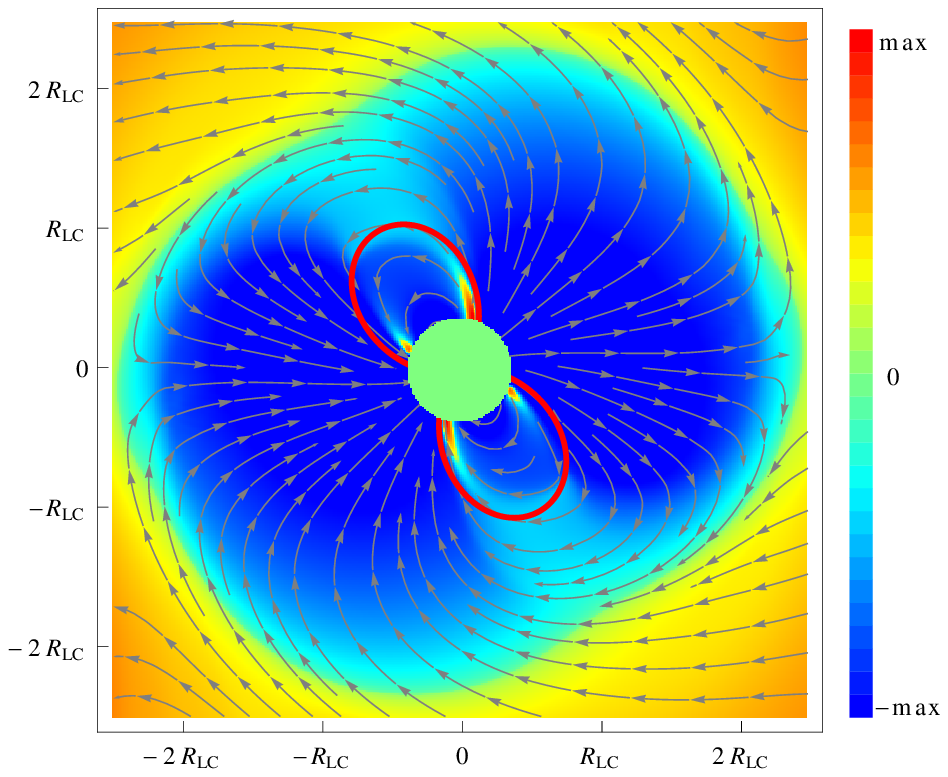}\hfill
\includegraphics[width=0.5\textwidth]{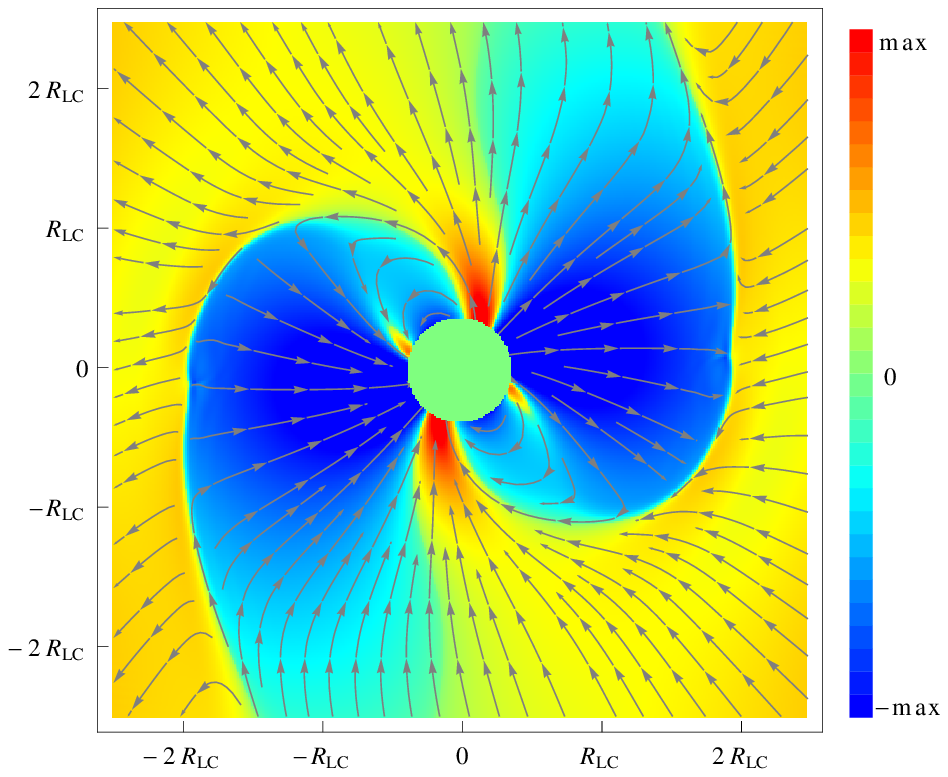}\\
\includegraphics[width=0.5\textwidth]{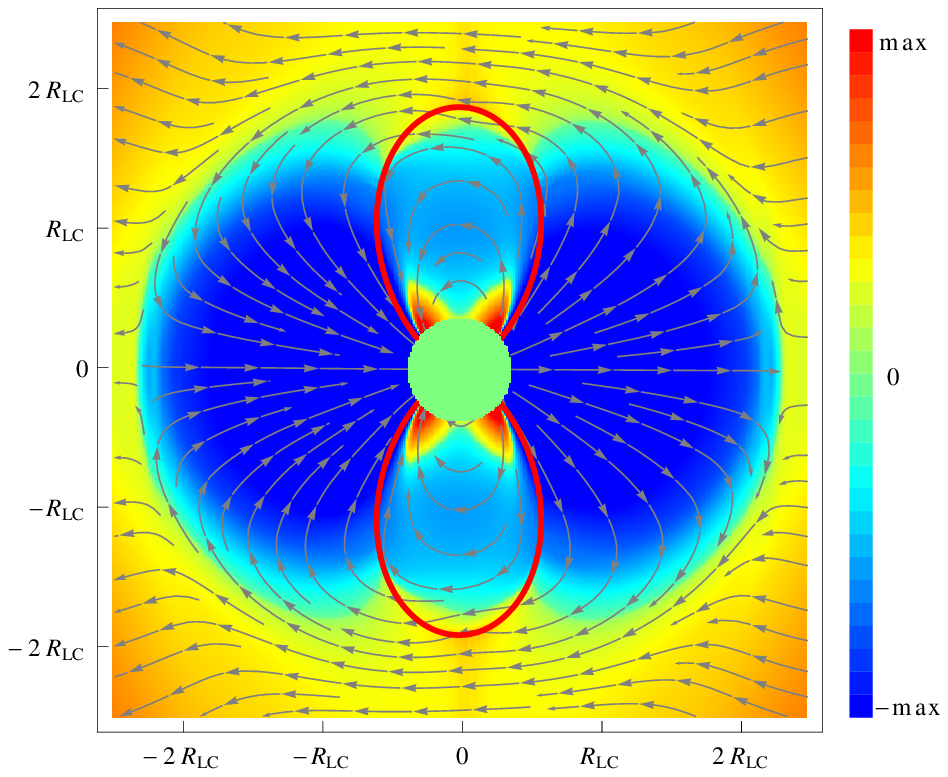}\hfill
\includegraphics[width=0.5\textwidth]{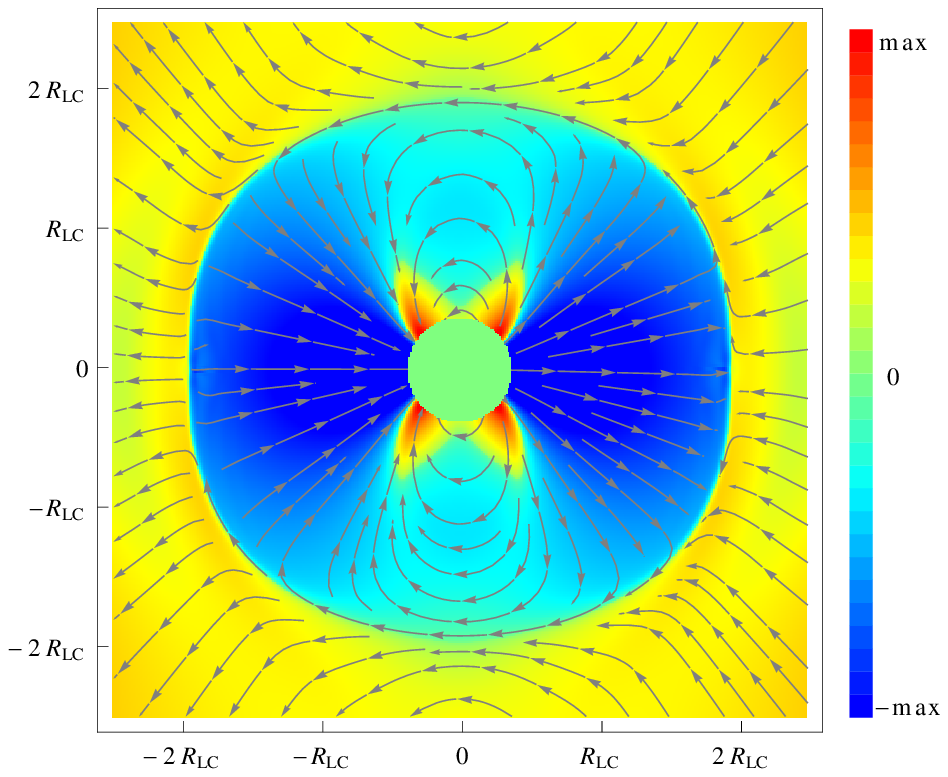}
\caption{Magnetic field lines in the $\vec{\mu}-\vec{\Omega}$ plane
for inclined dipoles at $\alpha=30^{\circ}$ (top), $\alpha=60^{\circ}$
(middle), $\alpha=90^{\circ}$ (bottom).  Color is representative of
out-of-plane magnetic field into (red) and out of (blue) the page.
The left column shows the ``off'' state with abundant plasma in the
closed zone and vacuum-like conditions along open field lines.  The
red curves indicate the intersection with the $\vec{\mu}-\vec{\Omega}$
plane of the demarcation surface between these distinct conducting and
vacuum-like regions.  Gross features of the magnetosphere in the
``off'' state are vacuum-like, but the ``off'' states have more
magnetic flux passing through the light cylinder and stronger magnetic
field sweepback than vacuum solutions.  The right column represents
the ``on'' state with abundant plasma everywhere.}\label{fig:fields}
\end{figure*}

\begin{figure}[t]
\centering
\includegraphics[scale=.8]{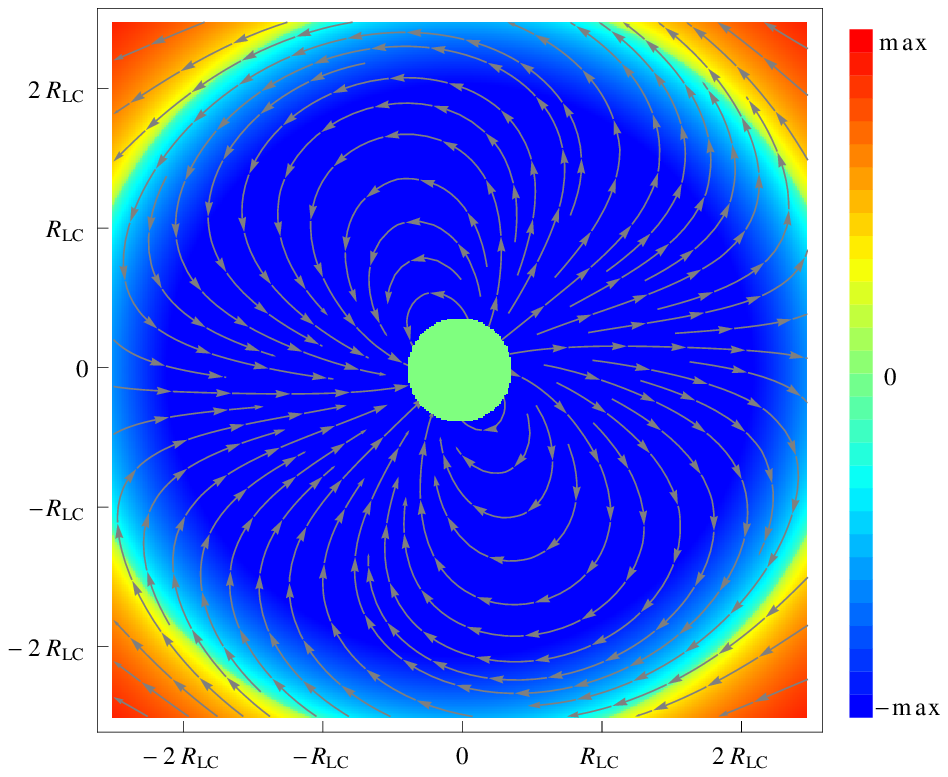}
\caption{Magnetic field lines in the $\vec{\mu}-\vec{\Omega}$ plane
for a vacuum inclined dipole at $\alpha=60^{\circ}$.  Color is
representative of out-of-plane magnetic field into (red) and out of
(blue) the page.  There is large closed field line region, and
displacement currents circulate and contribute to
spin-down.}\label{fig:vac}
\end{figure}

Figure \ref{fig:fields} shows magnetic field lines in the
$\vec{\mu}-\vec{\Omega}$ plane for inclined dipoles at
$\alpha=30^{\circ}$ (top), $60^{\circ}$ (middle), $90^{\circ}$
(bottom).  Color is representative of the out-of-plane magnetic field
(as in LST11).  The left column shows the ``off'' state with abundant
plasma in the closed zone but a shortage of plasma along open field
lines.  The red curves indicate the boundary between the conducting
and vacuum-like regions in the illustrated cross-section of the
magnetosphere.  The right column shows the ``on'' state with abundant
plasma everywhere.  The magnetospheres in the ``on'' state are
force-free--like, with conduction currents flowing along open field
lines.  The current returns through the current sheets and along the
boundary of the closed field line region in the current layers.  Gross
features of the magnetosphere in the ``off'' state are vacuum-like,
with large closed field line region and displacement currents
contributing to spin-down.  For reference we show a representative
vacuum solution with inclination angle $\alpha=60^{\circ}$ in Figure
\ref{fig:vac}.  There are a number of important differences that
distinguish the intermittent ``off'' state from the vacuum solutions.
The current from toroidal advection of charged plasma in the closed
zone leads to greater magnetic flux passing through the light cylinder
and a larger fraction of open field lines.  Further, poloidal
conduction currents are present even outside the conducting closed
zone.  They are due primarily to the fluid advection term
$\rho\vec{v}$ in the current and lead to greater magnetic field
sweepback and stronger current sheets than in pure vacuum solutions.

\begin{figure}[t]
\centering
\includegraphics[scale=.4]{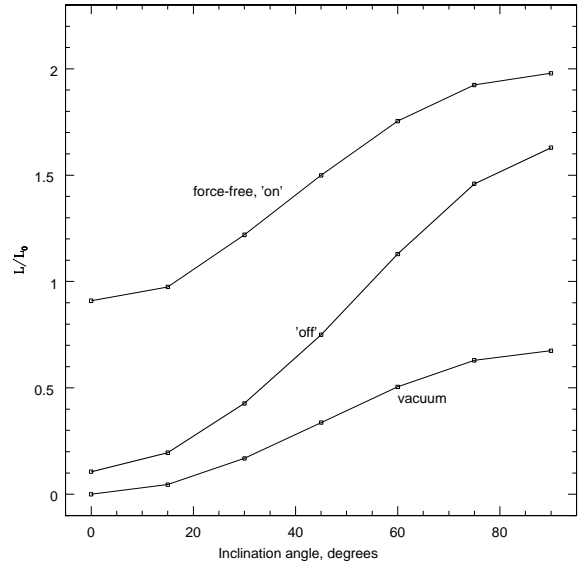}
\caption{Spin-down luminosity dependence on inclination angle for the
force-free--like ``on'' state, the ``off'' state, and the vacuum
solution.  The ``off'' state spin-down naturally lies between the
force-free and vacuum spin-down values for all inclination angles.
spin-down is normalized by $3/2$ times the spin-down power of the
orthogonal vacuum rotator.}\label{fig:spindown}
\end{figure}

Figure \ref{fig:spindown} shows the spin-down luminosity, normalized
by $L_0$ (where $L_0$ is $3/2$ times the power of the orthogonal
vacuum rotator with finite $R_*/R_{\rm LC}=3/8$), as a function of
inclination angle for both the ``on'' and ``off'' states of the
magnetosphere.  Spin-down luminosity is calculated as the surface
integral of the Poynting flux over a sphere of radius $R_{\rm LC}$.
We also show for reference the spin-down for the vacuum solution in
the total absence of plasma.  The ``on'' state spin-down is
essentially the force-free spin-down.  There is an uncertainty of
roughly $10$\% of $L_0$ in the force-free spin-down at low inclination
angles due to stellar boundary effects and unphysical dissipation
above the polar caps (see LST11).  The spin-down in the ``off'' state
lies between the force-free and vacuum spin-down values at all
inclination angles.  The spin-down of the aligned rotator in the
``off'' state is small, as the displacement currents are zero and
large-scale conduction currents are weak.  The open field lines carry
minimal Poynting flux.  Inclined dipole open field lines carry
Poynting flux, however, and the larger magnetic flux passing through
the light cylinder in the ``off'' state results in higher spin-down
than in the vacuum solution.  This increase in Poynting flux over
vacuum spin-down in the ``off'' state is largest at high inclination
angles, where the open field lines in the vacuum solutions carry the
most Poynting flux.  The advective poloidal conduction currents also
lead to higher spin-down since they cause larger magnetic field
sweepback than in vacuum solutions.  We evaluated the relative
contributions of these two effects to higher spin-down over vacuum
solutions by killing the advection current term $\rho\vec{v}$ outside
the closed zone.  The larger magnetic flux passing through the light
cylinder is responsible for upwards of two-thirds of the increase in
Poynting flux over vacuum spin-down in the ``off'' state.

The spin-down in the simulated ``off'' state is relatively insensitive
to the size of the conducting closed zone.  We varied the volume of
the conducting closed zone by a factor of a few while keeping the size
of the polar cap fixed, and we find that spin-down changes by less
than $20$\% of $L_0$ at all inclination angles.  It is not the total
size of the conducting closed zone, but rather its extent on the
stellar surface that is more important in determining the gross
magnetospheric properties.  Blocking radial currents near the stellar
surface limits the large-scale poloidal current circuit.  Spin-down
also depends very weakly on the fiducial values of $\sigma/\Omega$
that we pick to represent highly conducting and vacuum conditions.  We
checked this by setting the conductivity to $\sigma/\Omega=0$ instead
of $(\sigma/\Omega)^2=0.04$ along open field lines, and we find that
the resulting spin-down is offset by less than $20$\% of $L_0$ for all
inclination angles.  The aligned ``off'' state spin-down drops to zero
in this case, as we expect, since the large-scale conduction currents
have been eliminated.  We also tried setting the closed zone to have
conductivity $(\sigma/\Omega)^2=20$ instead of $(\sigma/\Omega)^2=40$,
and spin-down results change by less than $10$\% of $L_0$ at all
inclination angles.

\begin{figure}[t]
\centering
\includegraphics[scale=.4]{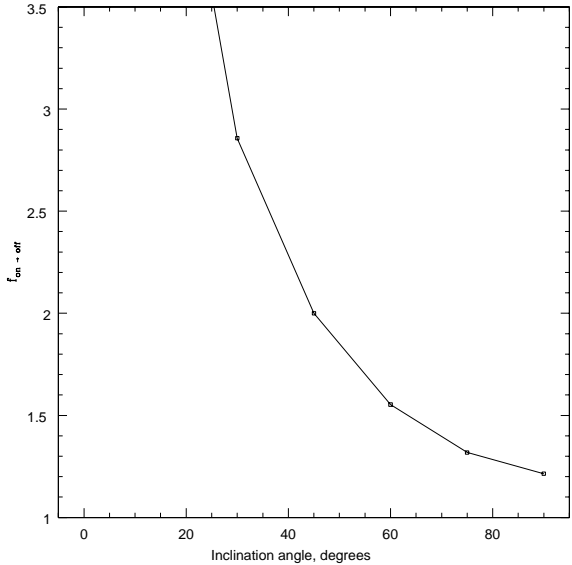}
\caption{The ratio of spin-down power between the ``on'' and ``off''
states, $f_{{\rm on}\to{\rm off}}$, as a function of inclination
angle.  We obtain $f_{{\rm on}\to{\rm off}}\sim 1.2-2.9$ for
$\alpha>30^{\circ}$.  Known intermittent pulsars have $f_{{\rm
on}\to{\rm off}}$ in this range.}\label{fig:ratio}
\end{figure}

The most relevant observational parameter in the context of
intermittent pulsars is the ratio of spin-down power in the ``on'' and
``off'' states, $f_{{\rm on}\to{\rm off}}$.  Figure \ref{fig:ratio}
shows this ratio for our models.  We emphasize that these models
provide a physically well-motivated set of solutions to describe both
the ``on'' and ``off'' states of intermittent pulsars.  Assuming a
uniform distribution of pulsar inclination angles, we expect
two-thirds of intermittent pulsars to have $f_{{\rm on}\to{\rm off}}$
between $1.2-2.9$.  Known intermittent pulsars fall within this range
\citep[K06,][]{Lyne09, Camilo11}.  Uncertainties in the exact charge
configuration and charge transport properties in the ``off'' state
magnetosphere imply error bars associated with our calculation of the
ratio $f_{{\rm on}\to{\rm off}}$.  The uncertainty is of order $10$\%
at inclination angle $\alpha=90^{\circ}$ and rises with decreasing
inclination angle to a factor of order $2$ at $\alpha=30^{\circ}$.
Below $\alpha=30^{\circ}$ the ratio $f_{{\rm on}\to{\rm off}}$ rises
above 3 due to the small spin-down of the aligned rotator in the
``off'' state, but there are large uncertainties in the exact value of
$f_{{\rm on}\to{\rm off}}$ in this regime.

A major improvement of our intermittent pulsar models over existing
work (LST11) lies in the treatment of accelerating potential drops,
which can be used as a fiducial measure of energy gain as particles
fly away from the stellar surface.  We define the potential drops as
in LST11, i.e., as the line integral along magnetic field lines of the
corotating electric field
$\vec{E}'=\vec{E}+(\vec{\Omega}\times\vec{r})\times\vec{B}/c$.  In
ideal force-free solutions the potential drops vanish, as
$\vec{E}\cdot\vec{B}=0$, but they increase monotonically with
increasing bulk resistivity.  Previously, we implemented a constant
conductivity $\sigma/\Omega$ throughout the magnetosphere.  As the
polar cap played no fundamental role in these models, open field line
potential drops were generally limited by the full pole-to-equator
potential drop, yielding unphysically large potential drops of order
$10^{16}$ V.  Our new models for the ``off'' state introduce
conducting plasma in the closed field line zone, effectively shielding
the potential drop there.  The potential drops are then limited by the
polar cap potential drop $V_{\rm pc}=|\vec{\mu}|/R_{\rm LC}^2$.  For
typical pulsars with periods $P\sim 1$s, we obtain characteristic
potential drop of $\sim 10^{12}$ V, more in line with expectations.

\section{Discussion}\label{sec:discussion}
We have developed an improved numerical method for solving pulsar
magnetospheres with resistivity, and we apply it to describe
intermittent pulsars in a self-consistent manner.  In the ``on''
state, plasma is abundant everywhere, and the instabilities in this
plasma generate coherent radio emission.  In the ``off'' state plasma
has leaked off the open field lines, suppressing the open field line
currents.  The radio emission is hence shut off.  Plasma remains
trapped in the closed zone, however, and the current from toroidal
advection of these charges leads to spin-down values that are a factor
of $\sim 2$ larger than vacuum values.  This allows us to naturally
produce spin-down ratios $f_{{\rm on}\to{\rm off}}\sim 1.2-2.9$ for
inclination angle in the range $\alpha=30^{\circ}-90^{\circ}$,
consistent with observations, and we obtain realistic values for
accelerating potential drops, $V_{\rm drop}\sim 10^{12}$ G.  In our
model the spin-down ratio $f_{{\rm on}\to{\rm off}}$ takes on its
minimum value at $\alpha=90^{\circ}$ and increases monotonically with
decreasing inclination angle.  Hence, given an observed spin-down
ratio $f_{{\rm on}\to{\rm off}}$ for an intermittent pulsar, we can
predict, within errors, the pulsar's inclination angle.
Alternatively, given the pulsar inclination angle from, e.g.,
polarization vector sweep data, we can predict the spin-down ratio
$f_{{\rm on}\to{\rm off}}$ if the pulsar displays clear ``on'' and
``off'' states.  A verification of our predictions would lend strong
support to the idea that abrupt changes in plasma supply on open field
lines can play an important role in determining the emission
properties and spin-down rates of real pulsars.

One potential limitation of our models relates to the spatial
distribution of conductivity in the magnetosphere.  We specify only a
conducting torus located in the plane perpendicular to the magnetic
axis.  In fact, rotation of a conducting neutron star leads to
unipolar induction, and domes of negative charge form above the
magnetic poles even in the absence of a pair cascade if the work
function of the surface is low
\citep{Krause84,Krause85,Petri02a,Petri02b,SpitkovskyArons02,Spitkovsky04}.
We explored this effect by prescribing additional conducting domes
above the magnetic poles, setting $(\sigma/\Omega)^2=40$ interior to
the surfaces specified by $s=R_{\rm LC}\cos^2\theta$.  Spin-down
values increase at all inclination angles by an offset of less than
$20$\% of $L_0$, but the spin-down ratio $f_{{\rm on}\to{\rm off}}$
remains within the errors quoted in Section \ref{sec:results}.

It is important to note that at present the precise mechanism by which
the plasma supply is turned on and off is still unclear.  In at least
one intermittent pulsar case the switching between different spin-down
rates is quasi-periodic (K06), implying that the processes supplying
and limiting plasma alternately recur.  Intermittency may actually be
the same basic process as nulling, the crucial difference being that
the ``off'' state lasts months to years instead of from a few rotation
periods up to days.  In this light, timing noise can in some instances
be caused by nulling events that are not resolved by the observations.
The pulsar timing data in the Jodrell Bank data archive are typically
smoothed and do not resolve features that occur on time scales shorter
than a few tens of days \citep[see][]{Hobbs10,Lyne10}.  Suppose
recurring nulling events last for an accumulated time that is of order
a few percent of the time that the pulsar behaves normally.  Since in
our model the spin-down in the ``off'' state is typically of order
$1/2$ that in the ``on'' state, the resulting spin-down luminosity in
a state with unresolved nullings could easily be modified by $\sim
1$\%.  \cite{Lyne10} established a connection between timing noise and
mode changing, when the pulse profile of the pulsar changes to a
different shape, but they do not see mode changing in all pulsars that
exhibit timing noise.  It is possible that some of these pulsars
undergo nulling events, which leads to variation in spin-down rate and
the observed timing noise.  The mode changing is presumably due to
changes in the magnetospheric configuration. One possibility is that
it is an intermediate state between our ``on'' and ``off'' states in
which some but not all of the plasma along open field lines has leaked
away.  These intermediate states could lead to a number of observed
pulsar features including jitter, subpulse drift, precursors, and
interpulses, though such ideas are at present still speculative.
Another possible cause for changing pulse profiles is the presence of
multiple components to the emission, e.g., the core and conal
components, and one or more components being modified or shut off by
changes in plasma supply.

If the ``on'' and ``off'' states of the pulsar can last anywhere from
a few periods to years, it is further conceivable that Rotating Radio
Transients (RRATs) are another manifestation of the process captured
by our model.  RRATs are rotating neutron stars that occasionally emit
pulses, usually isolated, but in a few instances in a string of
several \citep{McLaughlin06, Palliyaguru11}.  It has not been ruled
out that RRATs are an extreme form of nulling, and we envision RRATs
as pulsars that spend the majority of their time in our ``off'' state,
but turn ``on'' from time to time and emit pulses.  If future
observations are able to establish a more definitive link between the
seemingly disparate processes of intermittency, nulling, timing noise,
and RRATs, it would be an important step in our understanding of
pulsar physics.

AT acknowledges support by the Princeton Center for Theoretical
Science fellowship.  AS is supported by NSF grant AST-0807381 and NASA
grants NNX09AT95G and NNX10A039G.  The simulations presented in this
article were performed on computational resources supported by the
PICSciE-OIT High Performance Computing Center and Visualization
Laboratory.

\end{document}